\documentclass[aps,pra,twocolumn]{revtex4-1}
\usepackage[latin1]{inputenc}
\usepackage{bm}
\usepackage{array}
\usepackage{color}
\usepackage{graphicx}
\usepackage{slashbox}
\usepackage[pdftex,colorlinks=true,plainpages=false]{hyperref}
\usepackage{multirow}
\begin{document}
\title{Clusters in Intense XUV pulses: effects of cluster size on expansion dynamics and ionization}


\author{Edward Ackad, Nicolas Bigaouette, Kyle Briggs and Lora Ramunno}
\affiliation{Department of Physics, University of Ottawa, Ottawa, Ontario K1N 6N5, Canada}


\date{\today}

\begin{abstract}
We examine the effect of cluster size on the interaction of Ar$_{55}$-Ar$_{2057}$ with intense extreme ultraviolet (XUV) pulses, using a model we developed earlier that includes ionization via collisional excitation as an intermediate step. 
%
%
%
%
We find that the dynamics of these irradiated clusters is dominated by collisions.
Larger clusters are more highly collisional, produce higher charge states,
and do so more rapidly than smaller clusters.
Higher charge states produced via collisions are found to reduce the overall photon absorption, 
since charge states of Ar$^{2+}$ and higher are no longer photo-accessible. 
We call this mechanism \textit{collisionally reduced photoabsorption}, and
it decreases the effective cluster photoabsorption cross-section
by more than 30\% for Ar$_{55}$ and 45\% Ar$_{2057}$.
%
An investigation of the shell structure soon after the laser interaction
shows an almost uniformly charged core with a modestly charged outer shell which evolves to
a highly charged outer shell through collisions. This leads to the explosion of the outer positive shell
and a slow expansion of the core, as was observed in mixed clusters at shorter
wavelength \cite{xenon_cluster_shell_13nm}. 
The time evolution of the electron kinetic energy distribution begins 
as a (mostly) Maxwellian distribution. Larger clusters initially have higher temperature, but are overtaken by smaller temperature after the laser pulse. 
%
%
The electron velocity distribution of large clusters quickly become isotropic
while smaller clusters retain the inherent anisotropy created by photoionization.
Lastly, the total electron kinetic energy distribution is integrated
over the spacial profile of the laser and the log-normal distribution of cluster size
for comparison with a recent experiment \cite{PhysRevLett.100.133401}, and
good agreement is found.
\end{abstract}


\maketitle

\section{Introduction}

The interaction of atomic clusters with intense ultrafast laser pulses has been investigated over a range of wavelengths in recent years. 
Most of this work has been in the infrared (IR) regime \cite{RevModPhys.82.1793}, but as new very intense shorter wavelength sources have come online, 
this interaction has been investigated up to the soft-Xray \cite{PhysRevLett.94.023001,PhysRevLett.99.213002,PhysRevLett.95.063402,Xenon_clusters_13nm,PhysRevLett.92.143401,0953-4075-42-13-134019,xenon_cluster_shell_13nm,PhysRevLett.105.083005}.

Cluster interaction with intense laser pulses is highly wavelength dependent. In the IR, tunnel ionization and electron heating processes dominate and 
the subsequent plasma dynamics are driven by the laser field itself \cite{RevModPhys.82.1793,Lora_book}. This is not true at shorter wavelengths. 
Experiments on clusters in the vacuum ultraviolet (VUV) regime near 100 nm \cite{wabnitz_nature} observed unexpectedly high ionic charge states 
and a much larger energy absorption than predicted by existing models \cite{PhysRevLett.92.143401}. This sparked a concerted theoretical 
effort over the last eight years. Santra \textit{et al.} proposed an enhancement to inverse bremsstrahlung heating (IBH) due mainly to using self-consistent 
potentials for ions \cite{PhysRevLett.91.233401} and later plasma screening effects \cite{walters:043204}. Jungreuthmayer \textit{et al.}  proposed a 
many-body dielectric recombination scheme for the strongly-coupled plasma electrons in the cluster whereby the electrons could then be driven by the 
laser field to reionize \cite{mbr}. Siedschlag \textit{et al.} proposed allowing photoionization to occur above the classical potential barrier of the 
neighboring ion \cite{PhysRevLett.93.043402}. Ziaja \textit{et al.} incorporated many of these and found they all played various roles depending on the 
intensity \cite{PhysRevLett.102.205002}. It is clear from these works that clusters irradiated by short laser pulses with wavelengths in regime of single 
photon ionization represent a new and theoretically challenging area of physics.


Recent experiments around 40 nm have begun to probe the regime in which the photoelectron has a significant amount of kinetic energy and ions are 
photoaccessible \cite{murphy:203401,PhysRevLett.100.133401}. 
Contrary to the shorter wavelength regimes, 
the laser field coupled very weakly to the cluster electrons, producing a photoelectron 
spectrum with almost no signal above the atomic photoelectron energy in gas. The lower energy region showed that the cluster was charging and cooling 
subsequent photoelectrons down to lower energy. The clusters also produced high charge states well above what was detected in gas. This interaction 
was also found to depend on cluster size \cite{murphy:203401,PhysRevLett.100.133401}. Further, it was found that even small clusters emitted ions 
with very little kinetic energy (less than 30 eV) \cite{PhysRevLett.100.133401}. An attempt to address the charge transfer and explore the explosion 
dynamics was done using heterogeneous clusters \cite{ xenon_cluster_shell_13nm,Bostedt_review_FEL}. Many aspects of these experiments are not well 
understood even in gases \cite{Richter_JPB_review_Xe_90eV}.

The XUV regime offers unique opportunities in the study of the dynamics of finite systems such as clusters. Single photon ionization from the ground state of rare gas atoms is accessible, though inner shell ionization is not. In addition, unlike in the IR and VUV, the IBH of freed electrons is negligible, due to the very low quiver energy, decoupling cluster plasma dynamics from the laser field \cite{ZiajaVUV_XUV}.
However, electrons ionized from atoms have a significant amount of kinetic energy, enough to collisionally ionize neutral Argon.  
To date, there have only few theoretical works in this regime, however.
Bodstedt \textit{et al.} proposed a multistep model to explain the photoelectron spectrum of their experiment on small Argon clusters exposed to intense 32 nm laser pulses \cite{PhysRevLett.100.133401}.
Single photon ionization events were determined based on a Monte-Carlo model, and after each ionization step an electron is emitted from the cluster with an energy that depended upon the building positive space charge of the cluster.
Ziaja \textit{et al.} also examined the photoelectron emission spectrum, but used a kinetic Boltzmann equation technique that was then compared to molecular dynamics simulations results \cite{ZiajaVUV_XUV}. They found good agreement with the measured electron emission spectrum and showed that IBH was insignificant at 32 nm. Arbiter \textit{et al.} used a molecular dynamics and Monte Carlo technique to look at the electron emission spectrum at different intensities finding that at high intensity there are a significant number of thermal electrons emitted from the cluster \cite{PhysRevA.82.013201}.

Most recently, we showed the importance of two-step collisional ionization via an intermediate excited state \cite{AckadPRL}. This model allowed atom and ions to first be excited by an impact electron, which requires much less energy than ionization, and then ionized from the excited state. We called this model augmented collisional ionization (ACI) and were able to reproduce the highest charge state seen in Ref.~\cite{PhysRevLett.100.133401}.
This paper will expand upon our previous work to give a more detailed analysis of Argon cluster interaction with intense 32 nm laser pulses, including examining the cluster explosion dynamics and how these depend upon cluster size. Understanding cluster dynamics in the XUV regime, while interesting in its own right, may also inform planned single shot large molecule imaging experiments at even shorter wavelengths \cite{PhysRevA.71.013415,0953-4075-43-19-194015}.

The paper is organized as follows. In section \ref{theory} we describe the model and including details of how photoionization, collisional ionization and augmented collisional ionization (ACI) are implemented.  Section \ref{results} presents the results for Argon clusters of size 55, 147, 561 and 2057 atoms, with Section \ref{ions} concentrating on ions and Section \ref{electrons} on electrons. In Section \ref{EXP} we obtain the calculated photoelectron spectrum corresponding to the experimental results of Ref.~\cite{PhysRevLett.100.133401}, and find good agreement. Finally, Section~\ref{summary} gives a summary of all the results. Atomic units ($\hbar = e = $ m$_e =1$) are used throughout unless otherwise specified.


\section{Theory}\label{theory}


In this work we use hybrid approach to simulate the time evolution of Argon clusters in an XUV radiation field, where classical molecular dynamics is used for ion and electron motion and a Monte-Carlo scheme is used to determine ionization events. Ionization rates are determined by the quantum mechanical transition cross-section for the various processes. The simulation begins with a neutral cluster of atoms. Ionization events result in electrons being created within the code, and in an increment in the charge state of the the parent neutral or ion. Collisional excitation events are also permitted, whereby a neutral or ion may become excited; in this case no new electron is added to the simulation but the parent neutral or ion is set to be in an excited state, which determines its future ionization potential and ionization cross-section.


In order to mitigate numerical heating, the classical motion of particles was calculated assuming the particles were Gaussian distributions, instead of point particles, via the following smoothed potential
\begin{equation}
 \phi =  
\left\{ \begin{array}{lll}
   Q/r & & r \ge \sigma \\
   QB \exp{\left(-\frac{1}{2} \left(\frac{r}{\sigma} \right)^2\right) }  & & r < \sigma  
    \end{array} \right.
\end{equation}
where $r$ is the radial distance, $B$ is the maximum potential depth for a singly-charged ion, $Q$ is the charge, and $\sigma$ is an effective smoothing radius given by
\begin{equation}
 \sigma = \frac{Q}{B}\exp{\left(\frac{1}{2}\right)}.
\end{equation}
Note that $\phi$ is continuously differentiable at the transition point $r=\sigma$. 
To ensure energy conservation, the potential of an electron and a singly-charged ion are set to be equal in magnitude, 
and the potential depth of the multiply charged ions are integer multiples of that of the singly-charged ion's potential depth. Newly ionized electrons are created at the same location as their parent ion to avoid dipole heating.


Ionization in the cluster environment is modeled as an isolated system within a background cluster potential. We take the cluster potential as constant over the outer-most electron's wavefunction. Thus the cluster potential at the location of an ion or neutral is taken as the threshold for ionization for that ion or neutral. All ionization processes are then calculated with respect to this threshold. 
This allows for the use of atomic photoionization and collisional excitation/ionization cross-sections with only a small error due to the approximation \cite{AckadPRL}.


In the XUV, direct photoionization is the only way for the radiation field to deposit energy into the cluster. The single photon ionization probability is determined at each time-iteration for every neutral/ion, depending on its charge state and the photon flux. The photon flux is determined via the intensity, which is modulated both by the time profile of the pulse and by photon absorption. After each photoionization event the intensity of the pulse is decreased by one photon. This more accurately models low fluence pulses. The cross-section for the photoionization of neutral Argon was obtained from Ref.~\cite{Argon_photo_neutral}. Those of ionic Argon were obtained using Los Alamos Atomic Physics Codes \cite{gipper}. Multiphoton ionization is negligible at this photon fluence \cite{PhysRevLett.100.133401}.


Two channels are available for collisional ionization: a single-step transition from the ground state and a two-step transition through an intermediate excited state called augmented collisional ionization (ACI) \cite{AckadPRL}. 
The cross-section of all energetically accessible states (excitation and ionization) are combined to give the total cross-section for the occurrence of an event (whether excitation or ionization). If the impacting electron's trajectory is within this cross-section an event will take place. A Monte Carlo scheme is used to determine which type of event takes place based on the relative weights of the cross-section for each transition.
Single-step collisional ionization cross-sections were calculated using the semi-empirical Lotz formula and coefficients \cite{Lotz}. The Born plane-wave approximation \cite{cowan} was used for both excitation and ionization from an excited state.


The excited states considered consist of a subset of all possible excited states. Only single electron excited states were used and of those, only the lowest eight states with $l<3$ were implemented. This subset is the most important as it contains the lowest energy states. Including more single electron excited states adds to the total collisional cross-section, although states near the threshold require almost as much energy as ionization and are thus almost as infrequent. The energies of the excited states and the cross-sections were obtained using the Hartree-Fock implementation of the Cowan code \cite{cowan}.

\section{Results}\label{results}
In this section we illuminate the details of the laser cluster interaction and determine the effects of cluster size on key measures of the cluster dynamics. These include: ion charge state evolution and distribution, excited state evolution, the relative importance of the relevant ionization channels, and electron energies and charge transfer. Generally, the interaction dynamics proceeds as follows. After the first few direct photoionizations, the cluster builds up a positive space charge
such that at some point subsequent photoelectrons are prevented from escaping the cluster; this is described Ref.~\cite{PhysRevLett.100.133401} via their multistep ionization model. Our model further allows the cluster-bound photoelectrons to either release further electrons via collisional ionization or cause collisional excitation of neutrals or ions in the cluster during and after the laser pulse.

The parameters of the radiation field for each data set were $\lambda=32.8$ nm, $I=5\times10^{13}$ W/cm$^2$ and a full-width at half-maximum of 25 fs. Closed-shell icosahedral Argon clusters of 55 (2 shells), 147 (3 shells), 561 (5 shells) and 2057 (8 shells) were used.
These clusters were relaxed according to a Lennard-Jones potential for neutral Argon.
The results presented are for clusters assumed to be in the laser focus, and each plot is an ensemble average over many simulations, the number of which was chosen to ensure that more than $4\times10^{4}$ atoms are included in the average.

\subsection{Ions}\label{ions}
The charge states of the ions are a signature of the dynamics of intense laser cluster interaction.
Charge states higher than what is accessible through photoionization is an indication that there are cluster-driven processes that are not present in intense laser interaction with the gas phase of the material.
This occurs in clusters when the photon fluence is high enough to have at least a modest amount of photoionization within a single cluster, so that collisional processes become relevant. The parameter range of this work is such that the interaction is well above the collisional threshold.

\subsubsection{Charge state}\label{chgstate}
\begin{figure}
 \centering
  \includegraphics[scale=0.18]{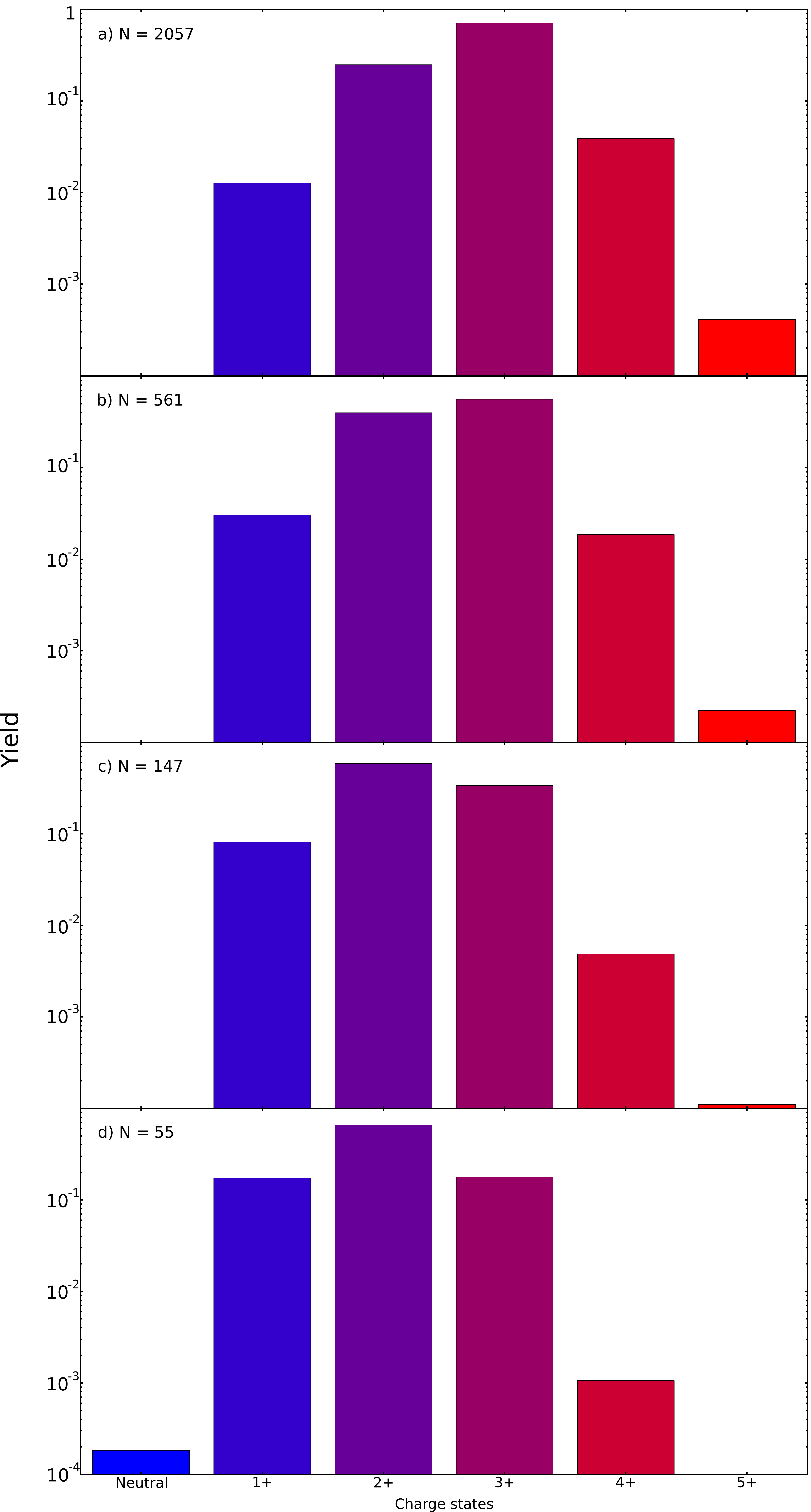}
 \caption{(color online)  The charge species yield per atom for: (a) Ar$_{2057}$, (b) Ar$_{561}$, (c) Ar$_{147}$ and (d) Ar$_{55}$ clusters interacting with a 25 fs, 32 nm laser pulse of 5$\times10^{13}$ W/cm$^2$. These snapshots were taken one picosecond after the pulse, after collisional events were no longer occurring.}\label{chgst}
\end{figure}

In Argon gas targets, only the Ar$^{2+}$ is accessible via direct photoionization. However experiments with clusters have detected higher charge states indicating the importance of cluster-driven processes such as collisions \cite{AckadPRL}. 
Figure~\ref{chgst} plots our calculated charge state distribution for Argon clusters of sizes (a) 2057, (b) 561, (c) 147 and (d) 55, for the laser parameters listed above.
%
Though photoionization stops after the laser pulse has passed, collisional ionization and excitation can still occur. Thus, these snapshots were taken after these processes also stopped, at one picosecond after the pulse. The bare charge states are reported.

We see that the average charge state increases with cluster size. This is despite the fact that larger clusters have fewer photons per atom compared with smaller clusters. However, larger clusters develop a larger space charge, and thus there are a greater 
number 
%
of photoelectrons that become bound to the cluster. These can then precipitate a collisional ionization or excitation event. Further, in a larger cluster there are more available collisional ionization or excitation targets. As we will see later in section \ref{mech_section}, collisional processes dominate over direct photoionization in charge state creation above Ar$^{1+}$.

\begin{figure}
 \centering
  \includegraphics[scale=0.36]{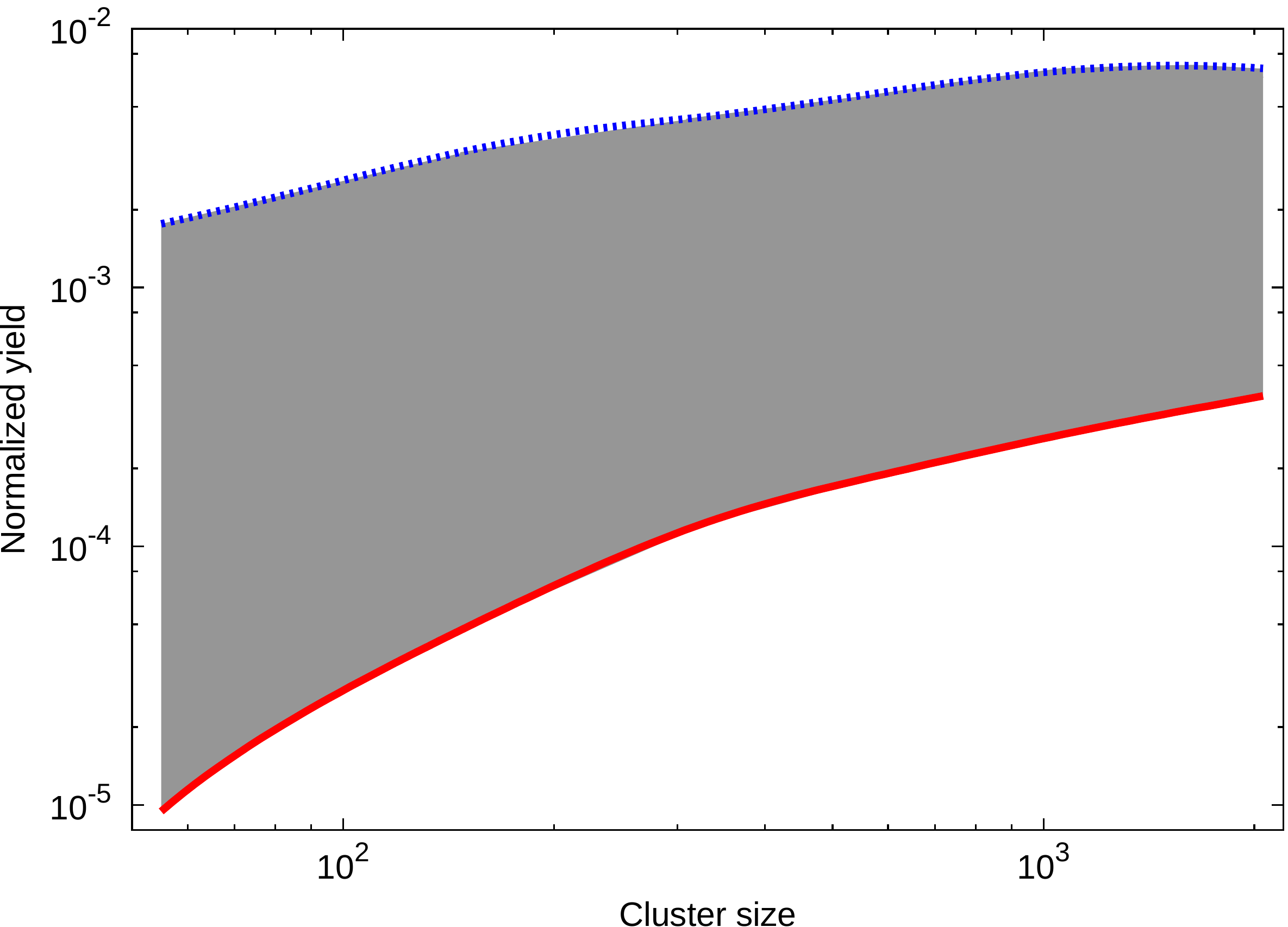}
 \caption{(color online)  The yield per atom of the Ar$^{3+}$ is shown by the (blue) dotted line and the Ar$^{4+}$ by the solid (red) line as a function of cluster size.
  The shaded (gray) region is only an aid to view how the difference between the lines diminishes with cluster size.
    }\label{4+vsN}
\end{figure}

Figure \ref{4+vsN} gives the yields per atom of Ar$^{3+}$ (blue dots) and Ar$^{4+}$ (red solid line) as a function of cluster size. Additional cluster sizes were considered for this graph.
The yields of both the Ar$^{3+}$ and Ar$^{4+}$ increase with cluster size, and moreover, 
the Ar$^{4+}$ yield increases more rapidly than the Ar$^{3+}$. This rapid increase of the Ar$^{4+}$ is thus not explained 
simply as an increase in Ar$^{3+}$, but indicates that collisional processes become more important for larger clusters. This is because it is only through collisional processes that charge states above Ar$^{2+}$ can be created.

\subsubsection{Charge state evolution}\label{chgstev}
\begin{figure}
 \centering
  \includegraphics[scale=0.32]{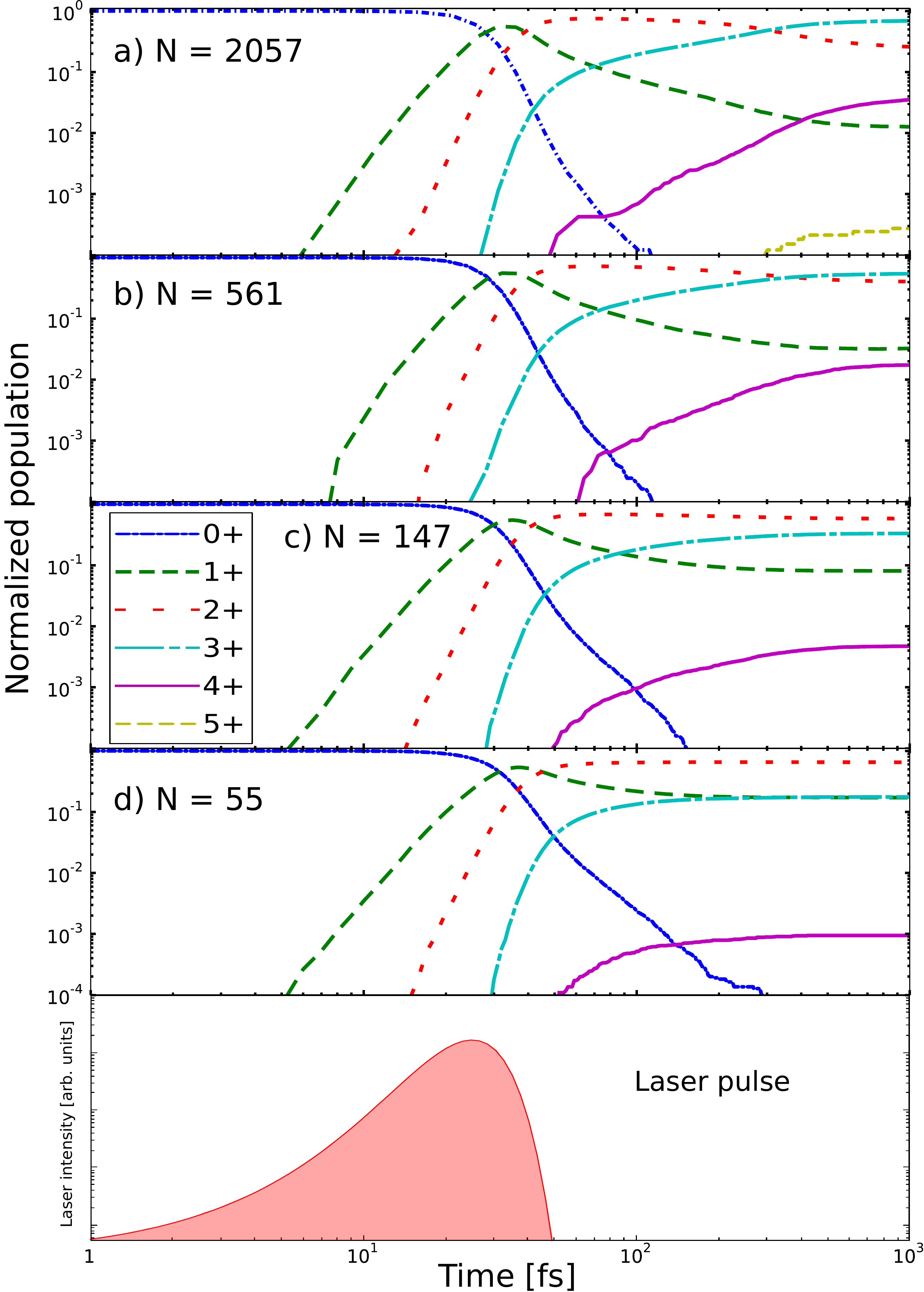}  
 \caption{(color online)   The charge species yield per atom as a function of time for: (a) Ar$_{2057}$, (b) Ar$_{561}$, (c) Ar$_{147}$ and (d) Ar$_{55}$ clusters interacting with a 25 fs laser pulse of 5$\times10^{13}$ W/cm$^2$. The bottom plot is the temporal profile of the 25 fs laser pulse.}\label{chgev}
\end{figure}

Our model also allows for the detailed tracking of each ion species during the interaction.
The normalized population of each charge state as a function of time is shown in Fig.~\ref{chgev} for different sized clusters.  Each plot shows the fraction of atoms that are neutral (blue dashed-dotted lines), Ar$^{1+}$ (green medium-dashed lines), Ar$^{2+}$(red sparsely-dashed lines),  Ar$^{3+}$ (cyan long and short-dashed lines), Ar$^{4+}$ (magenta solid lines) and Ar$^{5+}$ (yellow short-dashed line). At the bottom of the figure, we include a plot of the temporal profile of the Gaussian laser pulse. Note that a logarithmic scale is used for the time axis.

All clusters begin neutral and are initially ionized primarily by photoionization. The behavior of the low charge species early on is qualitatively similar for all cluster sizes. The population of neutrals decreases rapidly and is surpassed by the Ar$^{1+}$ followed by the Ar$^{2+}$. The Ar$^{2+}$ remains the most abundant charge state for the two smaller clusters. The larger clusters continue to have ionization well past 100 fs, and at around 300 fs the Ar$^{3+}$ becomes (and remains) the most abundant. After 1 ps, almost no further collisional ionization occurs.

The abundance of the higher charge states is quite small during the actual laser pulse. Contrary to longer wavelength regimes which are dominated by IBH, here photoionization is the only direct contribution of the laser pulse to charge state creation. The absorption of the XUV photons starts the ionization process, but the higher charge states appear in significant numbers only after the cluster has absorbed a sufficient amount of energy (as is further demonstrated in Section~\ref{eked_sec}). At the intensity we consider, this begins only near the tail end of the laser pulse and continues for several hundred femtoseconds.

The crossings of the curves in Fig.~\ref{chgev} occur earlier for larger clusters.
Given that there are less photons per atom for the larger clusters, this cannot be due to increased photoionization, and thus indicates that larger clusters are more efficient at dispersing the energy from the laser through increased collisions.

The location of the low charge state crossings all occur about 7 fs earlier for the Ar$_{2057}$ compared with the Ar$_{55}$.
This is almost 30\% of the full-width at half-maximum of the laser pulse, and results in a significant decrease in how much energy the larger cluster can absorb from the laser. This is because more ions become transparent to the laser earlier due to collisions in the larger cluster, and thus do not absorb as many photons. 
If photoionization were the dominant process, the low charge state 
crossings for larger clusters would occur later, not earlier, than for the smaller 
clusters, due to the larger number of photons per atom for the smaller clusters.
 

The fact that the neutrals decrease in population faster in larger clusters due to increased collisional ionization processes has other consequences. 
Neutrals have the largest photoionization cross-section. As their population decreases the overall photoionization cross-section of 
the cluster as a whole will also decrease, further decreasing the importance of photoionization in the larger clusters.
Further, photoionization increases the overall temperature of cluster-bound electrons, while it is decreased by collisional ionization. Thus, the 
electron temperature would be smaller when collisional ionization of neutrals is included because cluster-bound photoelectrons ionized from the neutrals contribute more energy than those photoionized from the Ar$^{1+}$. Thus a decrease in number of photoelectrons released from the neutrals would result in a decrease of the electron temperature.
Thus if these effects are neglected, this could overestimate energy absorption of the cluster.

We quantify the reduction of total cross-section and electron temperature due to collisions by performing our simulations with and without collisional processes included. Without collisional processes, we find Ar$_{2057}$ absorbs 1840 photons on average, 1401 by neutrals and 439 by Ar$^{1+}$. 
With collisional processes included, Ar$_{2057}$ absorbs 1151 photons, 618 by neutrals and 
533 by Ar$^{1+}$. Thus the total number of photons absorbed is decreased by 37\%, resulting in a 46\% reduction in the amount of energy the electrons obtain from the laser. We call this \textit{collisionally reduced photoabsorbtion} (CRP).


The effect is smaller in the less collisional smaller clusters. Without any collisional processes  Ar$_{55}$ 
absorbs 50 photons on average, 38 by neutrals and 12 by Ar$^{1+}$. With all collisional processes 
included Ar$_{55}$ absorbs 38 photons, 23 by neutrals and 15 by Ar$^{1+}$. Thus 
the total number of photons absorbed is decreased by 24\% in Ar$_{55}$ 
clusters. This still results in a 31\% reduction in the amount of energy the electrons obtain from the laser. While 
smaller than for the Ar$_{2057}$ clusters it is not a negligible effect.

\subsubsection{Excited states evolution}\label{ev_sec}

\begin{figure}
 \centering
  \includegraphics[scale=0.32]{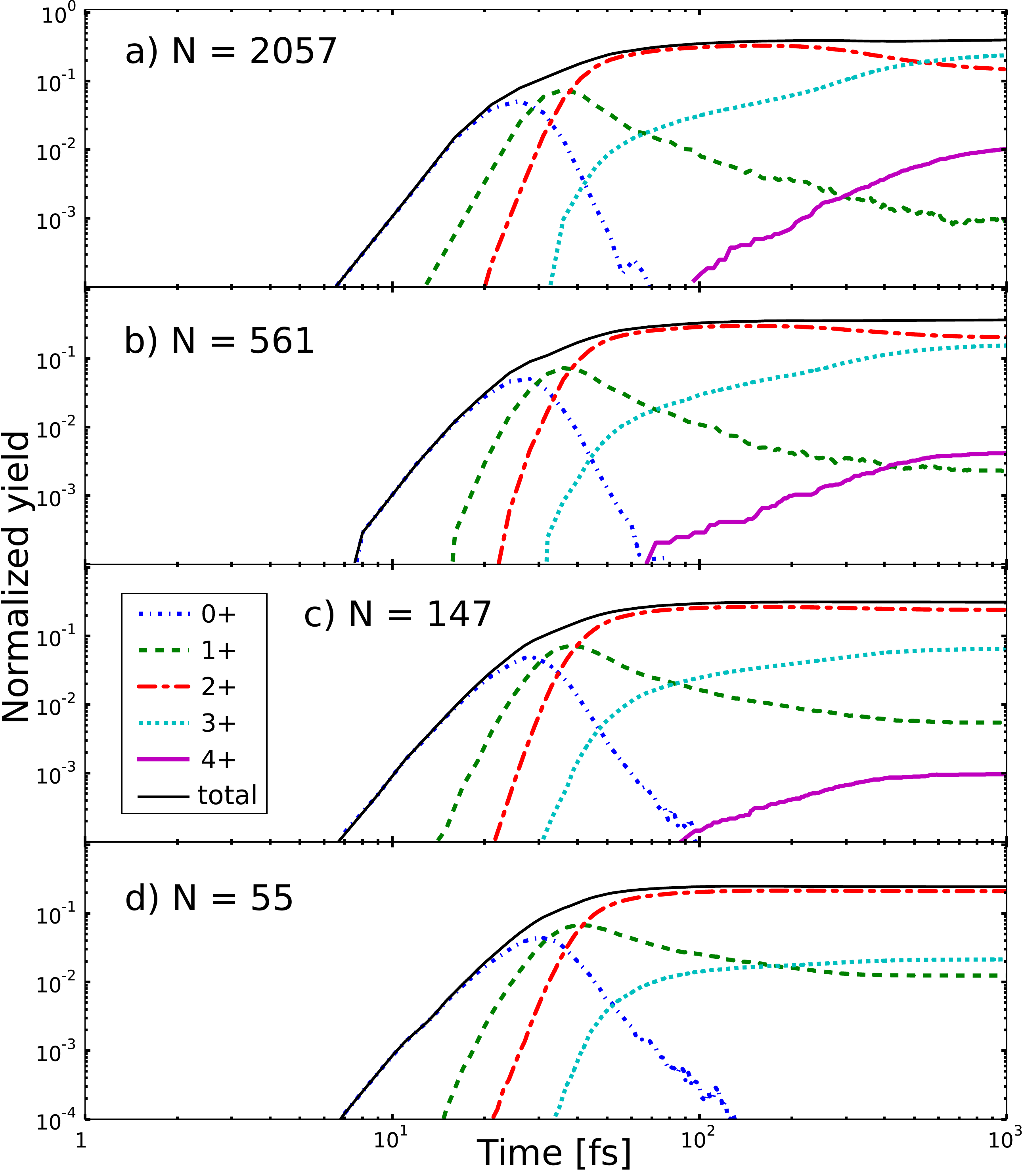}  
 \caption{(color online) Number of excited neutrals and ions, normalized to the total number of atoms, as a function of time for: (a) Ar$_{2057}$, (b) Ar$_{561}$, (c) Ar$_{147}$ and (d) Ar$_{55}$ clusters interacting with a 25 fs laser pulse of 5$\times10^{13}$ W/cm$^2$. The solid (black) line is the total number of excited neutrals and ions.}\label{esev}
\end{figure}

The efficacy of collisional ionization is driven largely by the access to excited intermediate states. Figure~\ref{esev} 
gives the excited state yield per atom of all neutrals and ions corresponding to the simulations of Fig.~\ref{chgev}. Also 
shown is the total number of excited ions as the solid (black) line. The curves are very similar in shape to Fig.~\ref{chgev}. 
Close to 20\% of all neutrals and ions for each charge species is excited at any given time. The proportion of excited species, 
for the most part, increases with the charge state. This is because the energy gap from an excited state to the threshold is larger 
for higher charge states. For lower charge states this energy gap is smaller, meaning that these excited species are comparatively 
shorter lived. The lower energy electrons are more abundant than higher energy ones, and thus excited states with a smaller gaps 
to threshold are more frequently ionized. The proportion of excited ions is higher at later times because collisions are much less frequent.

The evolution of the ionic and excited state populations show that the laser cluster interaction in the XUV is a process 
begun by photoionization but where collisional and photoionization processes occur in tandem, significantly affecting each 
other as quantified in Section~\ref{chgev}. The large amount of Ar$^{3+}$ in Fig.~\ref{chgev} and excited neutrals and ions shown in Fig.~\ref{esev}, where both snapshots are during the laser pulse, indicate that a significant amount of collisional ionization is occurring as these species can only be created via collisional processes. 
Further, the collisional processes  modify the photoabsorbtion rate 
of the cluster. More excited neutrals (as found in the larger clusters in particular) indicate that more neutrals could eventually be collisionally ionized via  augmented 
collisional ionization (ACI), which is the dominant cause of CRP as will be shown in the next section. 

Once sufficient energy has been deposited into the system by the laser, collisional processes (most notably ACI) disperse the energy throughout the cluster. The larger the cluster the more rapidly the cluster becomes dominated by collisional processes and thus reaches higher charge states.

\subsubsection{Mechanisms of ionization}\label{mech_section}

\begin{figure}
 \centering
 \includegraphics[scale=0.21]{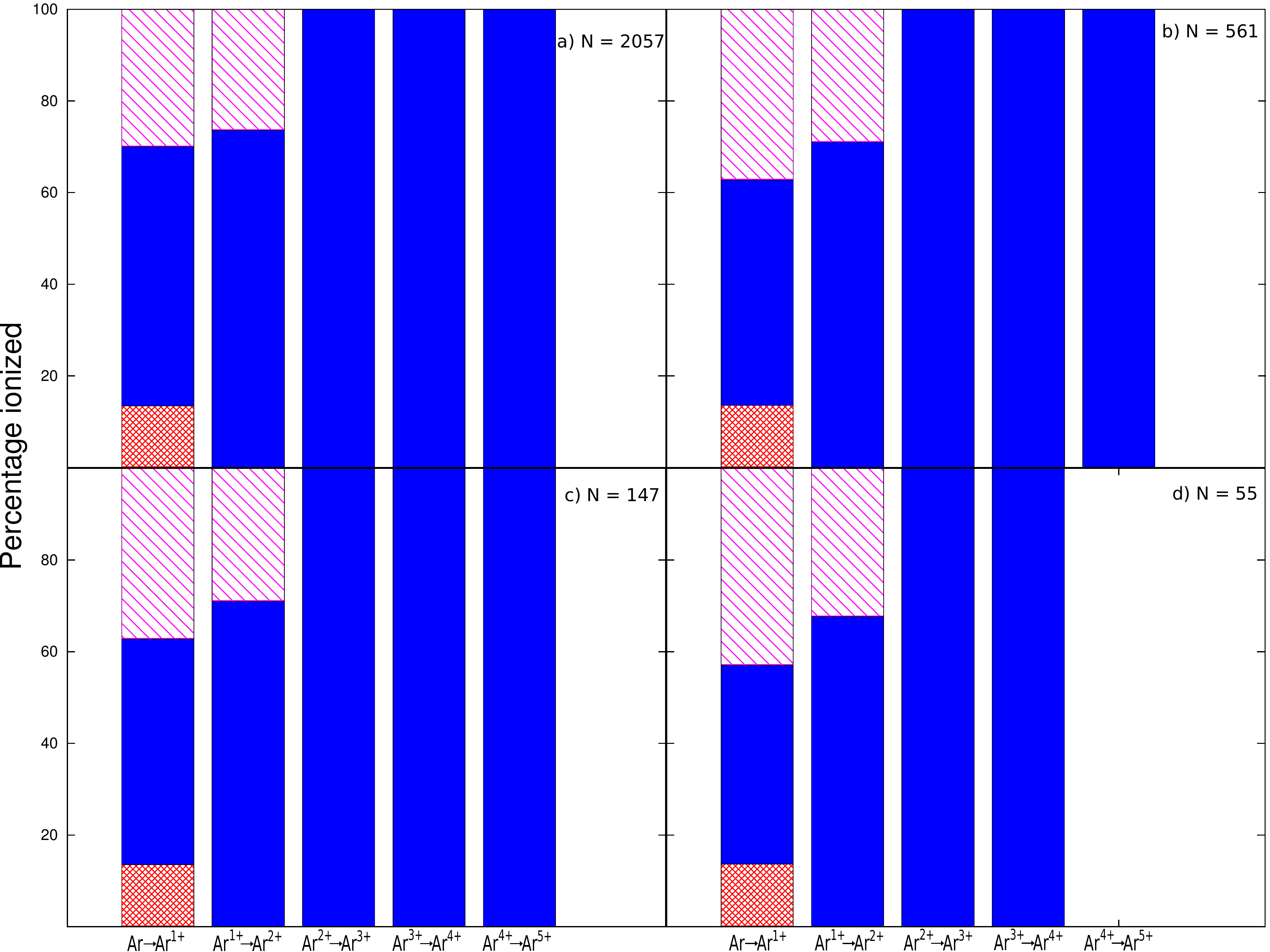}
 \caption{(color online)  Percentage of events that led to the ionization of the indicated charge state, for all the relevant mechanisms, for  (a) Ar$_{2057}$, (b) Ar$_{561}$, (c) Ar$_{147}$ and (d) Ar$_{55}$. The ionization mechanisms are: photoionization , single-step impact ionization  and augmented collisional ionization shown by (magenta) diagonal lines, (red) diamonds and solid (blue) fills respectively.}
 \label{mechanism}
\end{figure}

In order to examine the influence of each ionization mechanism directly, we plot for each charge state in Fig.~\ref{mechanism} the percentage of that charge state population that was ionized by the various available ionization mechanisms. The diagonal-lined boxes (magenta) give the percentage ionized by photoionization. The diamond-filled boxes (red) give the percentage ionized by standard one-step collisional ionization. The filled boxes (blue) give the percentage ionized via ACI.

For 32 nm radiation, only the neutral and Ar$^{1+}$ ions can be photoionized. For all cluster sizes, photoionization of neutrals is proportionally larger than for the Ar$^{1+}$, though it  
decreases for larger clusters, from 44\% for Ar$_{55}$ to 30\% to Ar$_{2057}$.
The photoionization of the Ar$^{1+}$ decreases from 36\% for Ar$_{55}$ to 28\% for Ar$_{2057}$.

The photoionization cross-section of Ar$^{1+}$ is close to half that of the neutral \cite{PhysRevLett.100.133401}, though this ratio is not born out in Figure~\ref{mechanism}. This is further evidence of how collisional processes affect the dynamics. We find that a significant number of neutrals are being ionized by collisional processes, predominantly ACI, before photoionization can occur. There is thus a change due to CRP in the expected photoionization yield for the cluster compared with what is expected for the same number of atoms in gas for the same photon fluence.

The proportion of neutrals ionized by single step collisional ionization is very similar for all cluster sizes at around 17\%. 
While it does in fact occur for higher charge states, it is below the 1\% range (thus not visible in the graph) demonstrating 
that ACI dominates. The proportion of neutrals and Ar$^{1+}$ ionized by ACI is larger for larger clusters. This accounts for the 
relative drop in the photoionization since the single-step collisional ionization remains roughly constant. This is due to the 
system becoming more collisional at earlier times for larger clusters increasing the effect of 
CRP.

The roles of photoionization and collisional processes are thus very clear. In this regime of intensity and wavelength, 
the system is initially driven by photoionization but its evolution is shaped by collisional processes. The higher 
charge states appear in large numbers well after the pulse is over, created by collisional processes which become more 
important for larger clusters.

\subsubsection{Charged shell structure}

\begin{figure}[t]
 \centering
\includegraphics[scale=0.408]{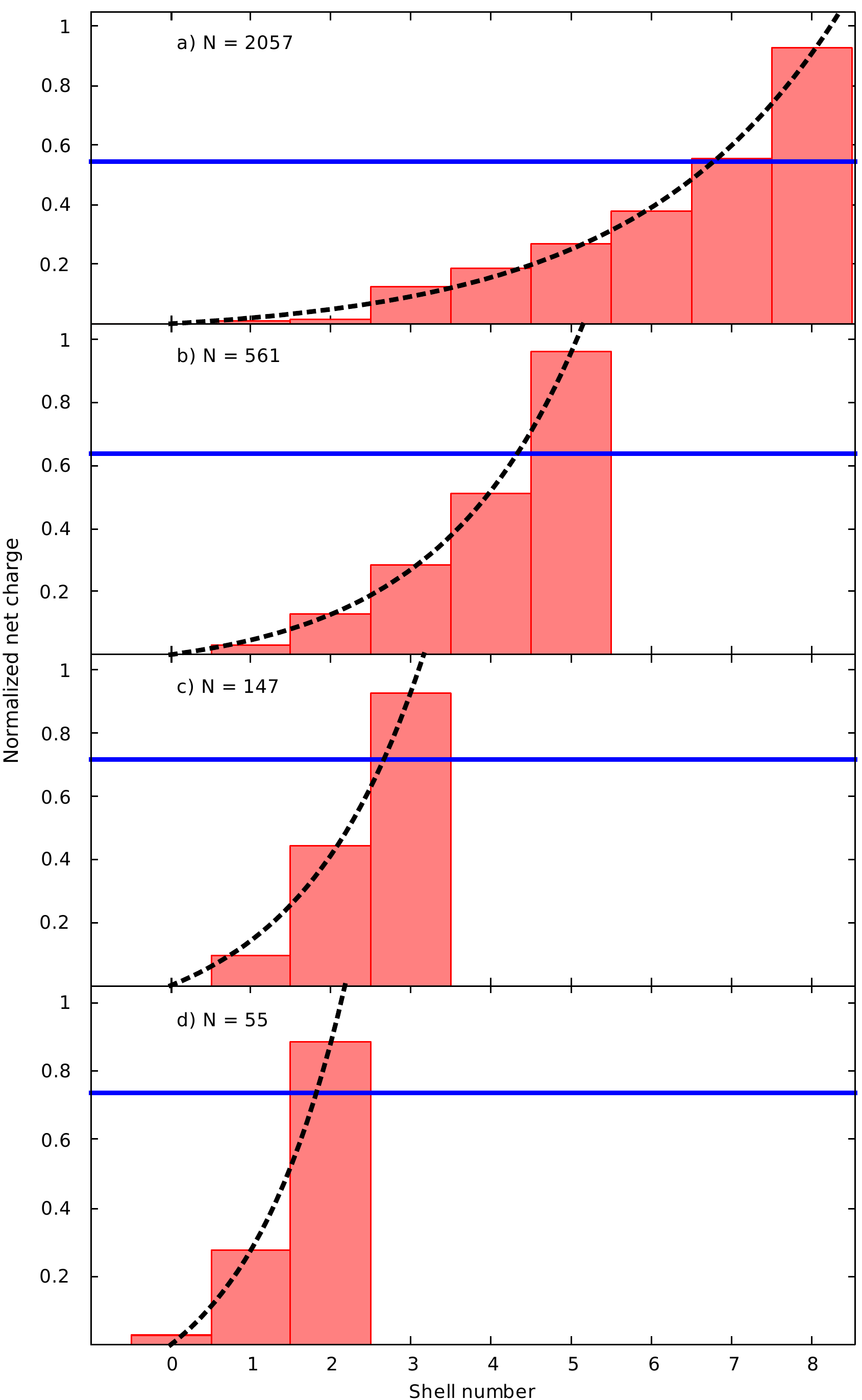}
 \caption{(color online)  Net charge per atom versus shell index for (a) Ar$_{2057}$, (b) Ar$_{561}$, (c) Ar$_{147}$  and (d) Ar$_{55}$ as determined at the end of the simulations. The solid (blue) line is the average of the net charge per atom over all the shells. }
 \label{shells}
\end{figure}

We now consider the spatial distribution of the charge states within the clusters. The initial icosahedral structure has closed shells which remain largely intact due to the force on the ions being primarily from their mutual repulsion \cite{Xenon_clusters_13nm}. Thus to understand the charge state distribution, we consider the net charge per atom as a function of shell index. The net charge within a shell is the total charge of all particles in that shell including ions and electrons classically bound to the ions; this is then divided by the number of atoms in the shell to obtain the net charge per atom. This is plotted versus shell index in Fig. \ref{shells} for Ar$_{2057}$, Ar$_{561}$,  Ar$_{147}$  and Ar$_{55}$ clusters at the end of the simulation, {\it i.e.}, after ionization processes have ceased.

The average over all shells of the net charge per atom is shown in as the horizontal blue solid line, and it gives a measure of how the cluster as a whole is charged.
It decreases with the increasing cluster size, indicating that while larger clusters access higher charge states the electrons remain bound to the cluster.
Consistent with previous work \cite{xenon_cluster_shell_13nm}, the net charge per atom of the outer shells is found for all cluster sizes to be much higher than that of inner shells. Moreover, only the outermost shell is above the average. Therefore in all cases most of the charge resides on the outermost layer(s) of the cluster. Taking the quasi-neutral core to consist of those shells which are below the mean, we find that the core size generally increases with increasing cluster size, leaving the outermost shell to explode fastest.

We can fit the net charge per atom versus shell number by an exponential function,
\begin{equation}
 f(s) = a(\exp(b(s-2))-1) ,
\label{fitfunction}
\end{equation}
where $s$ is the shell number, and $a$ and $b$ are fit parameters; $b$ quantifies the disparity in mean charge state across the shells. We plot these fits as black lines in Fig.~\ref{shells}, and list the fitting parameters in Table~\ref{fit_parms}.
The value of $b$ increases with decreasing cluster size, showing that smaller clusters have a larger disparity in the net charge per atom between shells, particularly between the outermost and next outermost shells.
Larger clusters have a smaller disparity, since the larger space charge retains more electrons overall.

\begin{table}
 {\begin{center}
\begin{tabular}[t]{|l|l|l|}
\hline
Cluster size & a & b\\  \hline
2057 & 0.041 & 0.394\\ \hline
561 & 0.062 & 0.562\\ \hline
147 & 0.157 & 0.646\\ \hline
55 & 0.230 & 0.789  \\
\hline
\end{tabular}
\caption{Fitting parameters for the net charge per atom versus shell index, according to the function given by Eq. \ref{fitfunction}.}\label{fit_parms}
\end{center}
}%
\end{table}

\begin{figure}[t]
 \centering
 \includegraphics[scale=0.45]{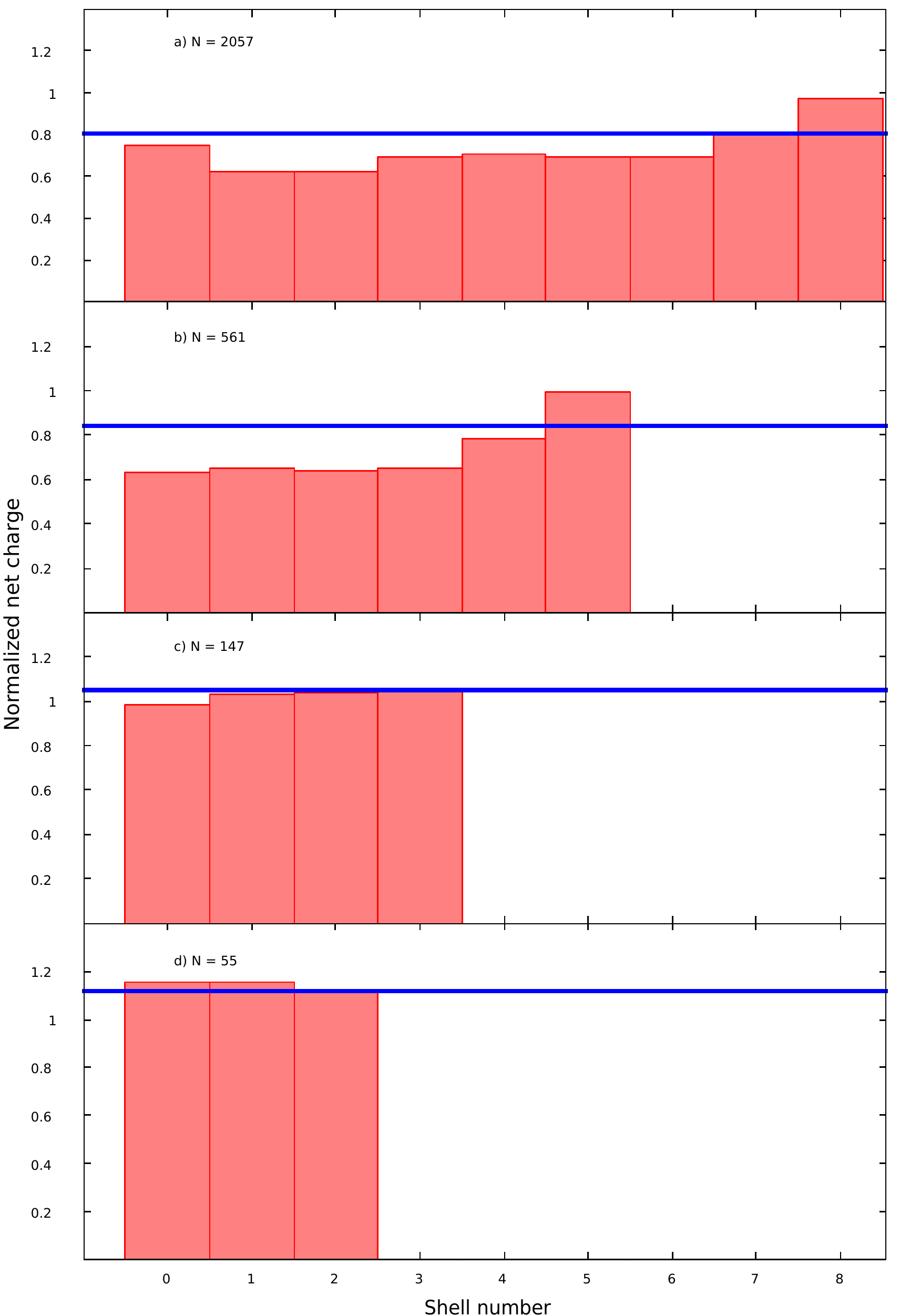}
 \caption{(color online)  Net charge per atom versus shell index for (a) Ar$_{2057}$, (b) Ar$_{561}$, (c) Ar$_{147}$  and (d) Ar$_{55}$ clusters after the laser pulse. The solid (blue) line is the average over all shells.
}
 \label{shellsafter}
\end{figure}

In Fig.~\ref{shellsafter} we plot a snapshot of the net charge per atom versus shell index shortly after the laser pulse (at 60 fs). 
The result is quite different from Fig.~\ref{shells}, which gives a snapshot long after the laser pulse when all collisional 
ionization processes have ceased. In  Fig.~\ref{shellsafter} the net charge per atom is more evenly distributed over the shells.
This implies that the shell structure is a consequence of charge migration through collisional ionization which is effectuated by the electrons.
For larger clusters, however, charged outer shells are already emerging, whereas for the Ar$_{55}$ and Ar$_{147}$ clusters they are still quite homogeneous.
This, along with Fig.~\ref{chgev}, suggests that the smaller clusters take longer to become collisional.
The formation of charged outer shells early on in the larger clusters is another indication of a significant overlap of the photoionization 
and collisional ionization dominated regimes as was found in Section~\ref{ev_sec}.

Figures~\ref{shells} and \ref{shellsafter} lead to the following explanation of the sequence of events. During the laser pulse 
the cluster charges up and some outer-ionization takes place. After the laser pulse collisional ionization causes charge migration 
from the outer shells to the core as electrons lose their energy by collisionally ionizing targets and falling deeper into the 
cluster's potential. The outer shells then explode faster than the inner shells which contain these cooled electrons.

\subsubsection{Kinetic energy\\}
\begin{figure}[t]
 \centering
 \includegraphics[scale=0.45]{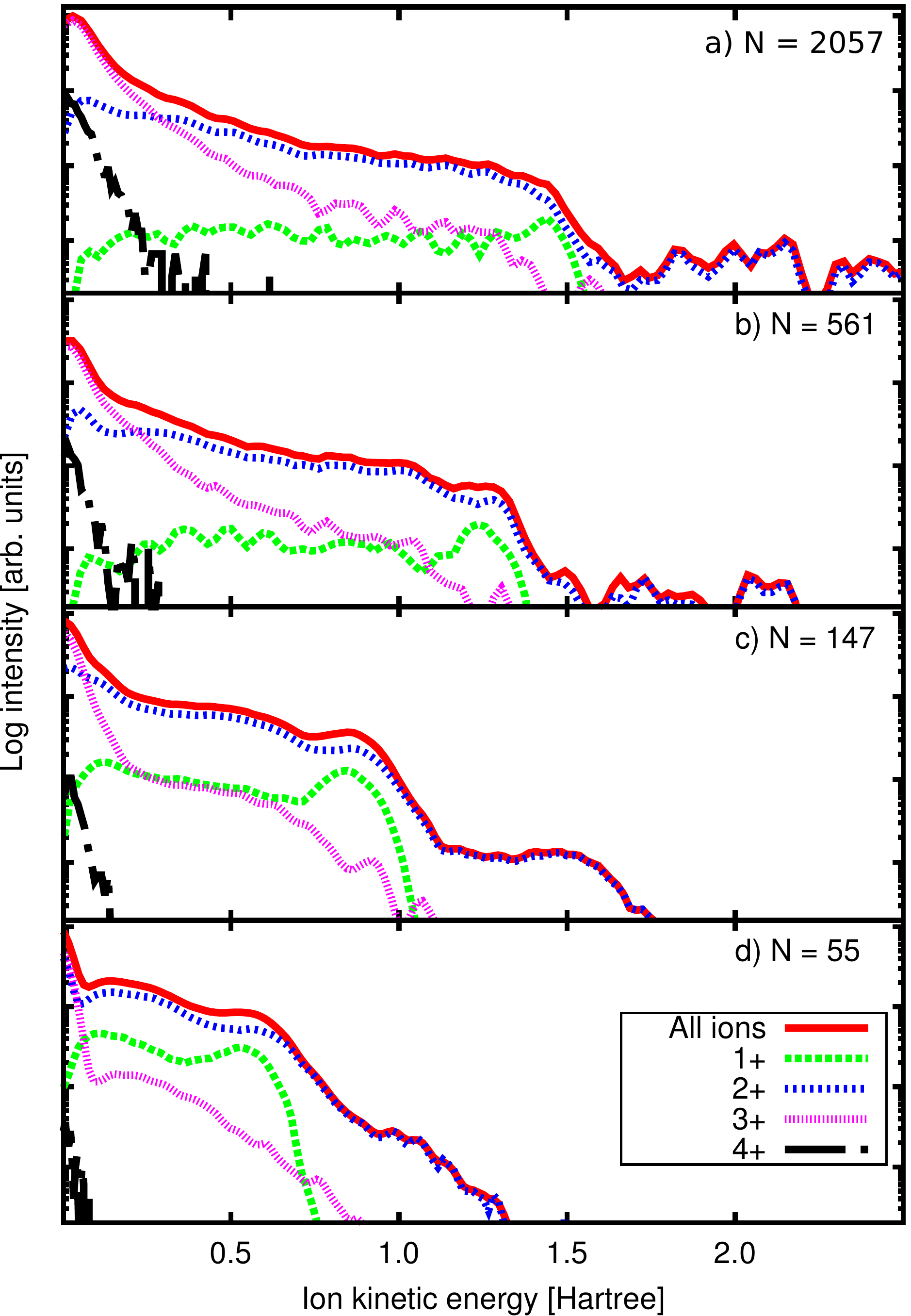}
 \caption{(color online)  Ion kinetic energy distributions of (a) Ar$_{2057}$, (b) Ar$_{561}$, (c) Ar$_{147}$  and (d) Ar$_{55}$ clusters at the end of the simulations for each charge state. The solid (red) lines is the total ion spectrum. }
 \label{ion_ke}
\end{figure}

The final kinetic energy distribution of the ions provides information about how the cluster has evolved after the laser pulse and is measurable in experiments.
In Fig.~\ref{ion_ke} we plot the ion kinetic energy distribution (red solid line) for Ar$_{2057}$, Ar$_{561}$,  Ar$_{147}$  and Ar$_{55}$ clusters at the end of the simulations. To gain further insight into how the cluster disintegrates, we also plot the kinetic energy distributions for each charge state individually: (green) long dashes for Ar$^{1+}$, (blue) medium dashes for Ar$^{2+}$, (magenta) short dashes for Ar$^{3+}$ and (black) dash-dots for Ar$^{4+}$.

The largest clusters produce the most energetic ions. For all cluster sizes the Ar$^{2+}$ is the most energetic and dominates the high energy end of the spectrum. For each cluster size, there is an obvious ``knee" in the total spectrum that clearly follows the Ar$^{2+}$ spectrum. This is because the Ar$^{2+}$ is the most populous ion in all but the inner most shells. Note that the shell location of the ions was determined by detailed positional analysis (not shown here) and is not evident from Fig.~\ref{ion_ke}. The Ar$^{2+}$ ions on the high energy side of the knee are from the exploding outermost shell and acquire a large amount of kinetic energy because there is a relatively low amount of electron screening. The Ar$^{2+}$ on the low energy side of the knee are from the other shells. 

Ar$^{4+}$ appear as the least energetic ions. This is due to most of the Ar$^{4+}$ are produced in the core where collisions are more frequent. Since the core is more shielded than the outer shells, these Ar$^{4+}$ are effectively screened and thus gain less kinetic energy.

The Ar$^{3+}$ populations decrease smoothly as a function of kinetic energy for the larger clusters, and dominate the total spectrum at low energies (below around 0.2 Hartree). There is a knee in the Ar$^{3+}$ spectrum for the smaller clusters, at 0.2 and 0.1 Hartree for Ar$_{147}$ and Ar$_{55}$, respectively, though the knee is somewhat washed out for the larger clusters. The ions with kinetic energy preceding the knee are in the core and expand slowly due to electron screening. The Ar$^{3+}$ dominates at low energy because it is the most populous ion in the core.

The Ar$^{1+}$ ions are distributed almost evenly over the range of energy for the larger clusters with only a small peak at the high-energy end. In the smaller clusters, this peak is more pronounced. In all cases, this peak arises from ions on the outer most shell of the clusters, with energies about half of the energy of the most energetic Ar$^{2+}$, indicating that the highest energy Ar$^{2+}$ and Ar$^{1+}$ are not screened. These high kinetic energy Ar$^{1+}$ (those at the peak) and Ar$^{2+}$ (those after the knee) primarily reside on the outermost shell and were ionized in the laser pulse solely by photoionization. They were thus not 
exposed to nor
screened by low energy electrons.
This explains why there is no similar phenomenon of a large knee or peak for Ar$^{3+}$ as they must be created by collisional processes.


In Fig.~\ref{ion_ke}c) a significant decrease in the Ar$^{1+}$ spectrum is evident around 1 Hartree. This is consistent with the experimental observations in Ref.~\cite{PhysRevLett.100.133401} for similarly sized clusters, which stated that Ar$^{1+}$ fragments were measured only up to 30 eV (1.1 Hartree). The largest contribution of high energy ions will come from the intensity peak of the laser pulse and from the largest clusters in the log-normal size distribution. The Ar$_{147}$ are thus the most likely origin of these fragments and our model agrees well with the experimental ion kinetic energy observed.

\subsection{Electrons}\label{electrons}
In the XUV, the laser interacts with the cluster through photoionization. The electrons are ejected from their parent 
ion or atom largely parallel to the laser polarization, but its subsequent motion is almost completely independent of 
the laser field due to the very small quiver energy at short wavelengths, even at these high intensities. Thus electron 
motion is determined entirely by the Coulomb fields of the other charged particles in the cluster. After photonionization, 
electrons that do not have enough energy to escape the cluster space charge will disperse its energy throughout the cluster via collisions.
The following sections provide details of how the energy is transferred, which requires a microscopic model of the relevant process to provide an accurate analysis of electron energy distribution.  

\subsubsection{Kinetic energy distribution}\label{eked_sec}
\begin{figure}[ht]
 \centering
 \includegraphics[scale=0.45]{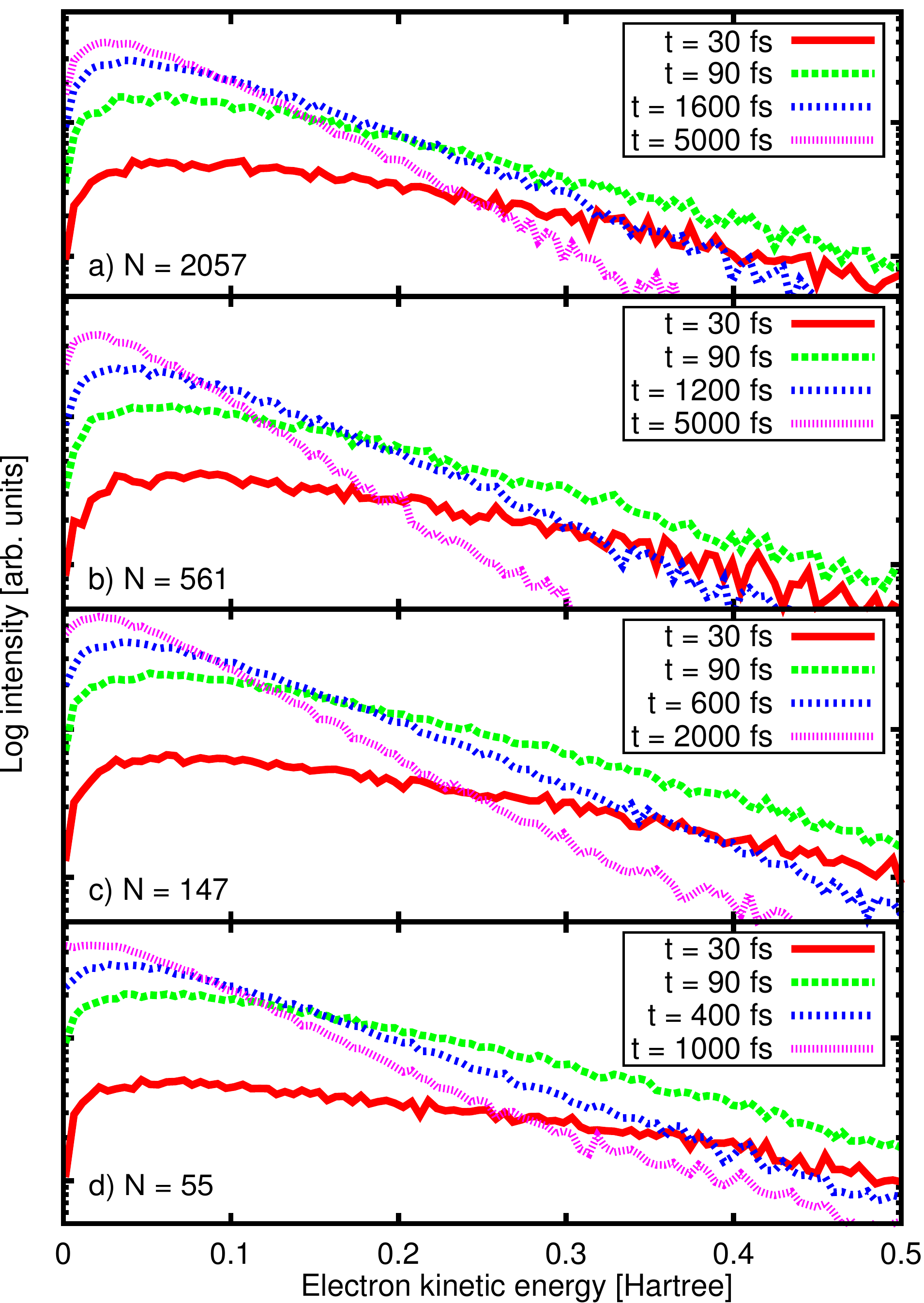}
 \caption{(color online)  Electron kinetic energy distribution for (a) Ar$_{2057}$, (b) Ar$_{561}$, (c) Ar$_{147}$  and (d) Ar$_{55}$. The time at which each distribution was calculated is indicated in the legends. }
 \label{EKED}
\end{figure}

The electron kinetic energy distribution for clusters of different sizes at different times is shown in Fig.~\ref{EKED}. These 
times are: near the peak of the laser pulse at 30 fs (red solid line), shortly after the laser pulse at 90 fs (green large-dashed line), 
after a significant decrease in the intensity of the high energy tail at times indicated in the legends (blue medium-dashed line), 
and at the end of the simulation when the electron distribution no longer changes, again at times indicated in the legends (magenta short-dashed lines).

The plots show that distributions at 30 fs are the only ones that deviate significantly from a single temperature Maxwellian 
distribution since the tails are not linear on the log plot. By 90 fs, they have become Maxwellian as indicated by the 
linear tails and this is verified by a quantitative analysis. Though not shown, 
the distributions of each cluster size remain the same for at least another 60 fs, before the clusters begin exploding. 
Thus the electrons thermalize quickly, indicating a high degree of collisions. Once the clusters start to explode at longer 
times, the distributions become less energetic, as evident from the graphs.

At 90 fs, there are proportionally more low energy electrons in larger clusters compared with smaller clusters, which have a proportionally larger number of high energy electrons; the differences are about $\sim$10 \% between the Ar$_{55}$ and Ar$_{2057}$, so may be difficult to see directly from the graph. This indicates that in larger clusters, fast electrons generally lose their kinetic energy more readily than in small clusters. This is consistent with the finding of Section~\ref{chgstev} that the larger clusters disperse the laser energy more quickly than small clusters. As there are more collisions in larger clusters, there will be faster thermalization and more collisional ionization, the latter of which actually removes energy from the ionized electron population. 

The aforementioned changes as the clusters evolve. At $\sim$150 fs (not shown) the distributions for all sizes are very similar, and at later times smaller clusters will have more lower energy electrons due to more rapid cluster explosion. Larger clusters have a larger quasi-neutral core and thus disintegrate more slowly; this allows for the preservation of more energetic electrons thus the distributions for the larger clusters take a longer time to become less energetic.
Note that the (blue) medium dashed line plots look very similar for each of the cluster sizes, but they were all calculated at different times, the latest time being for the largest cluster.


\subsubsection{Velocity distribution}
The laser polarization direction sets the axis along which most photoionization occurs. Anisotropic photoelectron emission was observed in low intensity synchrotron experiments, though emission from clusters was less anisotropic than from atoms 
\cite{PhysRevA.75.031201}. Only collisional processes can destroy the anisotropy inherent to photoionization. Since we have seen that clusters in intense laser pulses are highly collisional systems, which become more collisional as the size of the cluster is increased, we might expect that the effect of collisional processes would cause reduced anisotropy of the electron emission.

\begin{figure}
 \centering
 \includegraphics[scale=0.41]{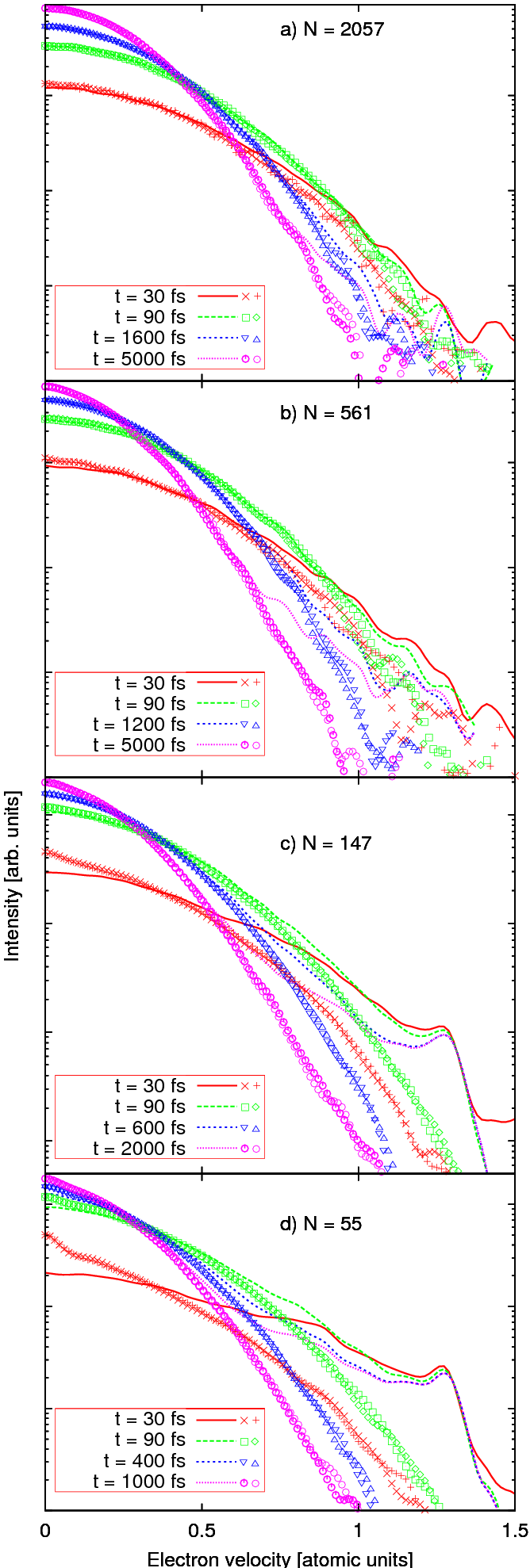}
 \caption{(color online)  Electron velocity distribution parallel (lines) and perpendicular (symbols) to the laser polarization for clusters of size (a) Ar$_{2057}$, (b) Ar$_{561}$, (c) Ar$_{147}$  and (d) Ar$_{55}$. The time at which each distribution was calculated is indicated in the key. Note that both perpendicular directions are considered, and these are both plotted for each time with similar symbols of the same color, as indicated in the key.}
 \label{EVD}
\end{figure}

Figure~\ref{EVD} shows the electron velocity distribution at different times for Ar$_{2057}$, Ar$_{561}$, Ar$_{147}$ and Ar$_{55}$ clusters. The times chosen are the same as in the previous section, and are also listed in the legends of the figures. The distributions parallel (lines) and perpendicular (symbols) to the laser polarization are shown for each cluster size and for each time. Though not shown explicitly, each distribution is symmetric about its respective axis. Data for both perpendicular directions are also included, though they are plotted with similar symbols of the same color for a given time. The degree of anisotropy is evident by the difference between the line and the points.

At 30 fs, near the peak of the laser pulse, the low velocity electrons are increasingly isotropic as cluster size increases. This indicates that larger clusters thermalize more rapidly than the smaller clusters, and are thus more collisional. 

There is increasing isotropy at high velocity as cluster size increases at all times. The high velocity electrons are those which have escaped the cluster and are no longer subject to collisions. In the small clusters, they thus form a peak centered at 1.27 Hartree, the 3$p$ photoelectron velocity.


It can also be seen that in Ar$_{147}$ clusters there is a larger decrease in intensity along the polarization axis from 30 fs to the final time at a velocity just below the peak, near 1 Hartree, than in Ar$_{55}$. This is due to the Ar$_{147}$ clusters having a larger space charge and being more collisional. This results in a decrease of the velocity of the later photoelectrons. The same is true in the larger clusters which are even more collisional and also have a larger space charge. Thus the Ar$_{2057}$ clusters have the largest difference as seen in Fig.~\ref{EVD}a. This trend explains the lack of a clear 3$p$ peak in the Ar$_{561}$ and Ar$_{2057}$ clusters.


The trend towards isotropy for larger clusters continues to increase for the highest velocities shown. The Ar$_{2057}$ clusters show the same amount of anisotropy at 30 fs as at the end of the simulation, indicating that
many of the early photoelectrons have undergone some collisions before exiting the cluster. 
In smaller clusters, even more electrons leave the cluster unabated. In the Ar$_{561}$, cluster this causes a larger anisotropy gap between the 30 fs and the end of the simulation than in Ar$_{2057}$. In the smallest clusters, this leads to the formation of the distinct photoelectron peaks.

The anisotropy at early times observed in the small clusters for low velocity electrons has similar origin. There are a significant proportion of electrons which exit the cluster unabated. This leaves only a few remaining electrons to collisionally ionize or excite neutrals or ions, which losing some of their kinetic energy and thus velocity. In larger clusters there are many more targets and a larger space charge at 30 fs which increases the number of collisions and removes the anisotropy at low velocity.

The electron velocity distribution shows that larger clusters are more collisional and become so earlier. This leads to an increase in the isotropy of the emitted electrons.

\subsubsection{Connection with experiment}\label{EXP}
Finally, we compare our simulations with experimental data of Ref.~\cite{PhysRevLett.100.133401} by calculating the electron energy spectrum over the intensity profile of the pulse. The spatial profile of the pulse was assumed to be Gaussian with a focus of 50 $\mu$m \cite{PhysRevLett.100.133401}, and the cluster jet size was taken to be 100 $\mu$m, the same size as the nozzle. The cluster size distribution was assumed to be log-normal with $\langle N \rangle= 80$ and $\Delta N=80$, which was estimated via simulations with Ar$_{80}$ as well as the two nearest closed-shell icosahedral clusters, ie. Ar$_{55}$ and Ar$_{147}$

\begin{figure}
 \centering
 \includegraphics[scale=0.37]{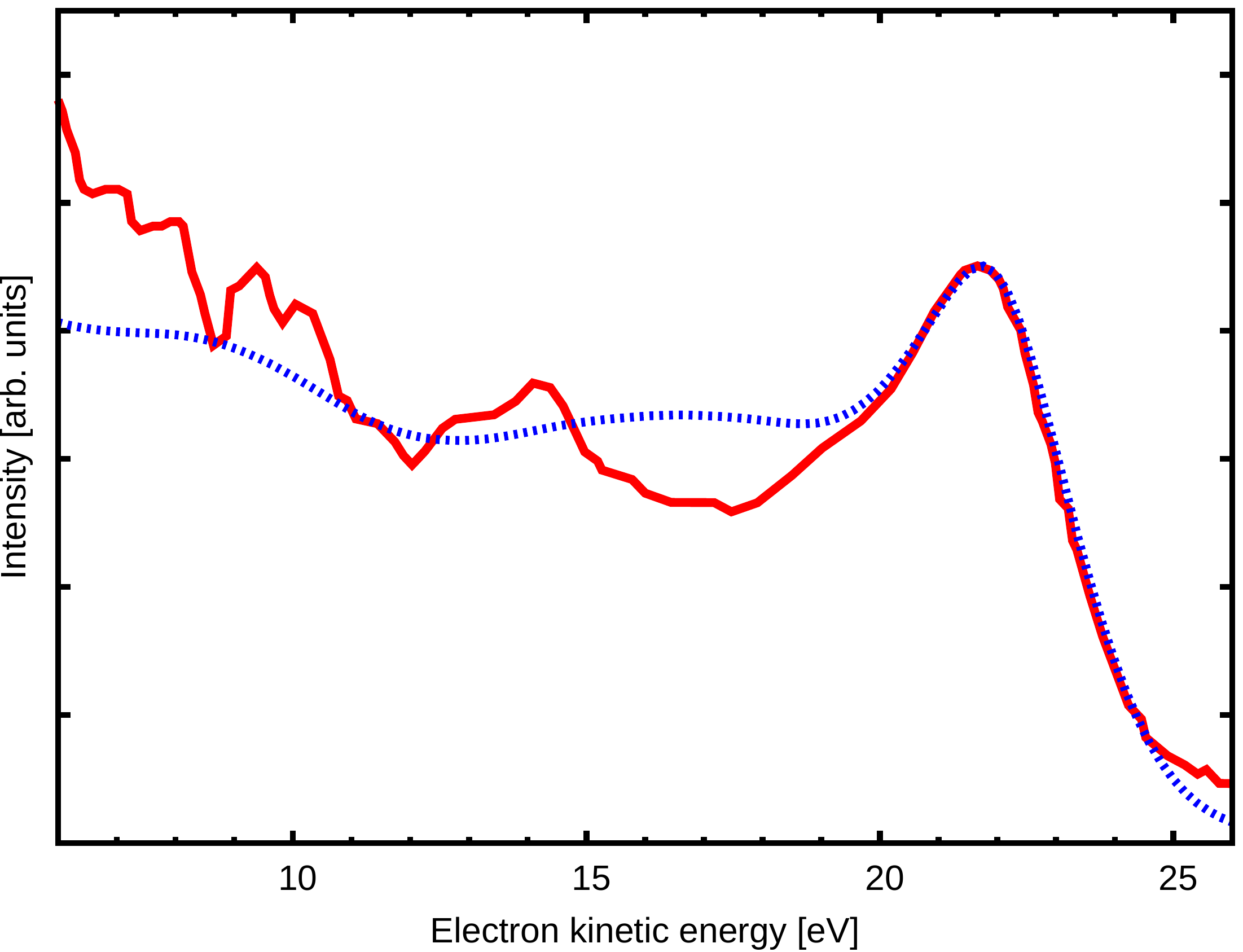}
 \caption{(color online)  The electron kinetic energy distribution integrated over the spacial profile of a Gaussian pulse with a peak intensity of $5\times10^{13}$ W/cm$^2$ a focus of 50 $\mu$m and integrated over a log normal distribution of cluster sizes with $\langle N \rangle= 80$ and $\Delta N=80$ is shown as the (blue) dashed line. The experimental data (red solid line) is taken from Ref.~\cite{PhysRevLett.100.133401}.}
 \label{espec}
\end{figure}

We calculated the electron energy spectrum by integrating over cluster size and laser intensity profile, for intensities from $5\times10^{13}$ W/cm$^2$ to $1\times10^{12}$ W/cm$^2$. Figure~\ref{espec} shows our result (blue dashed line) compared with the experimental result (red solid line) for a peak intensity of $5\times10^{13}$ W/cm$^2$ \cite{PhysRevLett.100.133401}. We found reasonable agreement, given our lack of knowledge of the laser profile and the precise experimental setup.

In Ref.~\cite{AckadPRL} the role of augmented collisional ionization was shown to be necessary in order to obtain the 
highest charge states seen in Ref.~\cite{PhysRevLett.100.133401}, Ar$^{4+}$. Figure~\ref{espec} shows that the model is 
also capable of explaining the dominant features in the electron spectrum seen in the experiment. The success of this 
model is due to the more accurate treatment ionization processes.

\section{Summary}\label{summary}
We have shown the effect of cluster size on many aspects of intense XUV-cluster interaction. 
Our model is verified by reasonable agreement with experimental observations in Ref.~\cite{PhysRevLett.100.133401}. This includes reproducing the electron emission spectrum, obtaining the highest observed charge state, and obtaining a close match of the 
maximum
kinetic energy of the Ar$^{1+}$ species for Ar$_{147}$ clusters.

We find that, for all measures and aspects of the cluster, collisional processes contribute significantly. Further, the size of the cluster increases the importance of the collisional processes. 
%
%
Larger clusters proceed from predominantly photoionization driven to collisional ionization driven more rapidly than smaller clusters. 
This causes a significant modification in the total cluster photoionization cross-section versus what would be expected 
from a gas, since less neutrals are photoionized when collisional processes are included. This results in a decrease in the deposition of energy by the laser. Even for a small cluster with 55 atoms, neglecting collisional processes will overestimate the amount of energy absorbed from the laser by over 30\%. We term this process \textit{collisionally reduced photoabsorbtion} (CRP).

The charge states were shown to be in greater abundance, proportionally, in the larger clusters. The highest charge states appear earlier in larger clusters and their appearance happens well into the disintegration of the cluster. At the photon fluence considered, almost no charge states above Ar$^{3+}$ were observed during the laser pulse. Those that were were created almost exclusively by a two step collisional ionization process, wherein an ion is first collisionally excited, then ionized from the excited state, termed augmented collisional ionization (ACI). ACI was shown to be the dominant collisional channel in general.  At any given time, around 20\% of neutrals and ions are in an excited state. 


 
An examination of the charge of the cluster shells as a function of time show evidence of charge migration from the outer to inner shells. Immediately following the laser pulse, the shells are almost uniformly charged, with the notable exception of the more highly charged outer shell in the larger clusters. However, as the system further evolves, collisions cause an increased charging of the outer shell of all cluster sizes, resulting the explosion of the outer shell and slow expansion of the inner shells, a process observed experimentally in mixed clusters \cite{xenon_cluster_shell_13nm}.


An analysis of the ion kinetic energy distribution of each charge state species found that the high energy tail was almost entirely due to Ar$^{2+}$. Most of the highest charge states were found to have very little kinetic energy. This because they are created in core of the cluster where they are shielded by electrons during the cluster disintegration.


The electron kinetic energy distribution was found to be close Maxwellian, except for very early times, indicating rapid electron thermalization. 
%
%
The electron velocity distribution was found to be largely isotropic for large clusters, but highly anisotropic for 
small clusters. This is due to the high velocity electrons in the small clusters originating from photoionization events. 
The isotropy in the large clusters is evidence of high velocity electrons undergoing multiple collisions before escaping the cluster.

Our findings may have implications for the direct single-shot imaging of large molecules with high intensity X-rays \cite{lcls_nature}. These systems will be highly collisional, which may, for example, lead to a rapid change in the photoabsorbtion cross-section of the molecule via CRP. Models of this interaction will need to account for this effect.

\begin{acknowledgments}
The authors would like to thank Thomas Brabec, Paul Corkum, Konstantin Popov and Jean-Paul Britcha for 
many fruitful discussions. This work is supported by the National Sciences and Engineering Research Council of Canada, the Ministry of Research and Innovation of Ontario, the Canada Research Chairs program, and the Canadian Foundation for Innovation. 
\end{acknowledgments}

\bibliography{expansion_dynamics}

\end{document}